\documentclass[aps,prl,showpacs,amsmath,amssymb,amsfonts,lengthcheck,longbibliography,superscriptaddress]{revtex4-1}
\usepackage{graphicx}
\usepackage{subfigure}
\usepackage{amsthm}
\usepackage{verbatim}
\usepackage{dcolumn}% Align table columns on decimal point
\usepackage{bm}% bold math
\usepackage{epsf}
\usepackage{color}
\usepackage[colorlinks=true,citecolor=blue,linkcolor=blue,urlcolor=blue]{hyperref}%
\usepackage{xcolor}
\usepackage{dsfont}

\newcommand{\tr}[1]{\mathrm{tr}\left\{#1\right\}}

\newcommand{\etal}{\textit{et al. }}
\newcommand{\e}[1]{\exp{\left(#1\right)}}

\newcommand{\bla}{bla\\bla\\bla\\bla\\bla}

\newcommand{\mc}[1]{\mathcal{#1}}

\newcommand{\be}{\begin{equation}}

\newcommand{\ee}{\end{equation}}

\newcommand{\ba}{\begin{align}}

\newcommand{\ea}{\end{align}}

\newcommand{\bi}{\begin{itemize}}

\newcommand{\ei}{\end{itemize}}

\newtheorem{theorem}{Theorem}

\begin{document}

\title{Negative entropy production rates in Drude-Sommerfeld metals}

\author{Marcus V. S. Bonan\c{c}a}
\email{mbonanca@ifi.unicamp.br}
\affiliation{Instituto de F\'isica `Gleb Wataghin', Universidade Estadual de Campinas, 13083-859, Campinas, S\~{a}o Paulo, Brazil}

\author{Pierre Naz\'e}
\email{p.naze@ifi.unicamp.br}
\affiliation{Instituto de F\'isica `Gleb Wataghin', Universidade Estadual de Campinas, 13083-859, Campinas, S\~{a}o Paulo, Brazil}

\author{Sebastian Deffner}
\email{deffner@umbc.edu}
\affiliation{Department of Physics, University of Maryland, Baltimore County, Baltimore, MD 21250, USA}
\affiliation{Instituto de F\'isica `Gleb Wataghin', Universidade Estadual de Campinas, 13083-859, Campinas, S\~{a}o Paulo, Brazil}

\date{\today}

\begin{abstract}
It is a commonly accepted creed that in typical situations the rate of entropy production is  non-negative. We show that this assertion is not entirely correct if a time-dependent, external perturbation is not compensated by a rapid enough decay of the response function. This is demonstrated for three variants of the Drude model to describe electrical conduction in noble metals, namely the classical free electron gas, the Drude-Sommerfeld model, and the Extended Drude-Sommerfeld model. The analysis is concluded with a discussion of potential experimental verifications and ramifications of negative entropy production rates. 
\end{abstract}

\maketitle

Most systems in the observable universe exhibit a tendency to evolve towards a state of equilibrium. Phenomenologically, this universal behavior is described by the non-negativity of the entropy production \cite{Callen1985}. More specifically, the thermodynamic entropy is \emph{defined} such that (i) it is maximal in equilibrium \cite{Callen1985}, and that (ii) all real, nonequilibrium processes are accompanied by the production of additional \emph{irreversible} entropy \cite{prigogine1961}. In irreversible thermodynamics it is then often	 taken for granted that also the \emph{rate} of entropy production has to be a non-negative function \cite{prigogine1961,mazur2011}, even to the extend that negative rates would signify the end of the universe \cite{Prigogine1986}.

Over the last couple of decades, the non-negativity of the entropy production rate has then become somewhat synonymous to a statement of the second law of thermodynamics  \cite{schnakenberg1976,mcadory1977,spohn1978,lebowitz1978,alicki1979,cohen1995,ruelle1996,seifert2005,deffner2019book}. However, in its original treatment \cite{prigogine1961} it is clear that the non-negativity is a direct consequence of the assumption of local equilibrium and, in particular, the rapid decay of the system's response to external perturbations. Thus, if either the rate of driving becomes comparable to one over the local relaxation time, or if the interaction between system and environment conspires to give rise to more intricate time-dependent response functions, finding negative entropy production rates is not that uncommon \cite{bylicka2016, Belandria2005,bhattacharya2017, marcantoni2017, popovic2018, xu2018, strasberg2019,naze2019}. See also a very recent review on the topic \cite{Landi2020}.

The natural question arises, whether there are any experimentally relevant scenarios in which negative entropy production rates are prevalent. Indeed, Williams \etal found \cite{Williams2007} that in viscoelastic fluids sheared by oscillatory forces, the entropy production rate has marked negative periods, while the average entropy production remains positive. In viscoelastic media the relaxation times are rather long, while at the same time the viscosity is high. Therefore, these negative rates are not a consequence of driving far from equilibrium, but rather an inherent property of the medium even in the regime of time-dependent linear response \cite{kubo2012,Williams2008}.

While viscoelastic fluids under oscillatory shears are interesting, they may not be the most common physical system. Therefore, the present letter is dedicated to the analysis of the entropy production rate in a much more mundane situation, namely noble metals under AC driving.  To this end, we  analyze the entropy production rates in three different scenarios starting with the classical Drude model \cite{Drude1900I,Drude1900II}, over the Drude-Sommerfeld model \cite{Sommerfeld1928}, and finishing with the Extended Drude-Sommerfeld model \cite{Allen1977,Beach1977,Youn2007}. The Extended Drude-Sommerfeld model is a sophisticated generalization of the classical free electron gas, which describes experimental findings to very high accuracy \cite{Olmon2012,Yang2015,Maslov2016}.

Specifically, we compute the entropy production rate with a standard tool of condensed matter theory, namely by means of time-dependent linear response theory \cite{mori1956, nakano1956,Ashcroft1976,giuliani2005,Williams2008,girvin2019}. We find that for all three cases, ranging from the simplest classical treatment to the sophisticated quantum model, negative periods of the entropy production rate are prevalent, while the entropy production itself remains strictly positive. In particular, we argue that these negative rates are a direct consequence of Ohm's law and thus fully consistent with phenomenological, irreversible thermodynamics. The analysis is concluded with a discussion of experimental consequences that could be observed in measurements of the conductivity in noble metals, such as silver \cite{Yang2015} and gold \cite{Olmon2012}.

%%%%%%%%%%%%%%%%%%%%%%%%%%%%%%%%%%%%%%%%%%%%%%%%%%%%%%%
\paragraph*{Preliminaries}

 In condensed matter theory \cite{Ashcroft1976,giuliani2005,girvin2019}, in particular in the description of electrical conduction of simple metals, it has been established that tools from linear, irreversible thermodynamics are apt and convenient. Thus, the main object of the present analysis is the entropy production rate, $\dot{\Sigma}$, written as a bilinear form, $\dot{\Sigma}=\sum_{i}\mathcal{F}_{i}(t)\mathcal{J}_{i}(t)$, of forces $\mathcal{F}_{i}$ and flows or currents $\mathcal{J}_{i}$ \cite{Prigogine1986,mazur2011}. The currents are given by the most general linear expression and we write \cite{Ashcroft1976,kubo2012}
\begin{equation}
\dot{\Sigma} = \sum_{i,j} \mathcal{F}_{i}(t) \,\int_{-\infty}^{t}dt'\,\Phi_{i j}(t-t')\, \mathcal{F}_{j}(t')\,,
\label{eq:eprdelay}
\end{equation} 
where $\Phi_{i j}(t)$ denote the response function. Note that for slowly-varying forces the \emph{time-averaged} response function becomes identical to the Onsager transport matrix \cite{Callen1985,kubo2012,Proesmans2015}. This is the ``usual case'', in which also the average entropy production rate remains strictly positive \cite{Maes2000,Bauer2016,Brandner2016}.  In the following, we will focus on the more general case of time-dependent response functions, $\Phi_{i j}(t)$, and we will see that $\dot{\Sigma}$ can, indeed, take negative values if the temporal variation of the forces $\mathcal{F}_i(t)$ cannot be compensated quickly enough by the decay of $\Phi_{i j}(t)$.

Equation~\eqref{eq:eprdelay} is not only at the core of the description of material properties \cite{Ashcroft1976}, but has recently also attracted significant interest in the study of thermodynamic control \cite{Sivak2012,zulkowski2012,bonanca2014,mandal2016,deffner2017,Bonanca2018,bonanca2019,scandi2019,Deffner2020EPL}. 
In this context it has been argued that the sign of $\dot{\Sigma}$ is rooted in whether or not the response function gives rise to a Riemannian metric \cite{Deffner2020EPL}.

In the present analysis, we focus on a physical scenario of immediate practical relevance, namely the entropy production in noble metals under AC electrical driving. To this end, we now proceed to compute the response function $\Phi(t)$ for three variants of the Drude model and determine the resulting entropy production rate. We start with the simplest case, and build up to more sophisticated scenarios. Despite its shortcomings and simplicity \cite{Ashcroft1976}, the Drude model does describe the properties of real metals such as gold, copper and silver \cite{Olmon2012,Yang2015} at room temperatures and low photon energies reasonably well.

%%%%%%%%%%%%%%%%%%%%%%%%%%%%%%%%%%%%%%%%%%%%%%%%%%%%%%%%%%%%%%%%%%%%
\paragraph{Classical Drude model}

In its original formulation \cite{Drude1900II,Drude1900II}, the Drude model describes a free electron gas of $N$ non-interacting and independent charge carriers. Nevertheless, it is assumed that these particles have a finite mean free path, which can be translated into a relaxation time. This quantity can be phenomenologically introduced expressing the collision term of the Boltzmann equation as \cite{Ashcroft1976}
\begin{equation}
\left(\frac{\partial f}{\partial t}\right)_{coll} = -\frac{\delta f}{\tau_{R}}\,,
\label{eq:coll}
\end{equation}
where $f$, $\delta f = f-f_{0}$ and $\tau_{R}$ denote the non-equilibrium distribution, the deviation from the initial equilibrium distribution $f_{0}$ and the relaxation time $\tau_{R}$, respectively. 

The relation between current, $\hat{\mathcal{J}}(\omega)$, and electric field, $\hat{E}(\omega)$, in frequency domain is given by Ohm's law \cite{Ashcroft1976},
\begin{equation}
\hat{\mathcal{J}}(\omega) = \sigma_{\mathrm{cl}}(\omega) \hat{E}(\omega)\,,
\label{eq:ohmslaw}
\end{equation}
where the conductivity $\sigma_{\mathrm{cl}}(\omega)$ reads
\begin{equation}
\sigma_{\mathrm{cl}}(\omega) = \frac{\sigma_{\mathrm{cl}}(0)}{1-i\omega\,\tau_{R}}\,,
\label{eq:classconduc}
\end{equation}
with $\sigma_{\mathrm{cl}}(0)=N q^{2} \tau_{R}/m$. Hence, Eq.~\eqref{eq:classconduc} reduces to the classical DC conductivity in the zero-frequency limit. 

Computing the inverse Fourier transform of Eq.~\eqref{eq:ohmslaw}, we obtain
\begin{equation}
\mathcal{J}(t) = \int_{-\infty}^{t}dt'\,\Phi(t-t')\,E(t')\,,
\end{equation}
where the response function $\Phi(t)$ is given by \cite{kubo2012}
\begin{equation}
\Phi_{\mathrm{cl}}(t) = \Theta(t)\, \frac{\sigma_{\mathrm{cl}}(0)}{\tau_{R}}\e{-|t-t'|/\tau_{R}}\,,
\label{eq:respclass}
\end{equation}
with $\Theta(t)$ denoting Heaviside step function.

Comparing the response function $\Phi(t)$ in Eq.~\eqref{eq:respclass} with the analysis of sheared viscoelastic fluids \cite{Williams2007}, we immediately conclude that the entropy production rate \eqref{eq:eprdelay} can indeed exhibit negative values for oscillating electric fields. We emphasize again that our approach is purely phenomenological as required by irreversible thermodynamics. Hence, the occurrence of these negative entropy production rates are fully consistent with the second law.

%%%%%%%%%%%%%%%%%%%%%%%%%%%%%%%%%%%%%%%%%%%%%%%%%%%%%
\paragraph*{Drude-Sommerfeld model}

Furthermore, a free electron gas can be understood as a charged ideal gas, for which also negative rates have been reported \cite{Belandria2005}. Yet, the classical Drude model is only a poor conceptual description of electrical conduction in real metals. Thus, we need to address the question whether these negative rates persist in more sophisticated models.  The Drude-Sommerfeld model is a quantum generalization of the classical free electron gas. In this model the classical, thermal distribution is upgraded to the Fermi-Dirac statistics \cite{Sommerfeld1928}. Nevertheless, the expression for the conductivity \eqref{eq:classconduc} remains formally the same. The quantum effects are encoded in modified values of $N$, $\tau_{R}$ and $m$. The number of charge carriers $N$ is given by the number of available electrons near the Fermi surface, $\tau_{R}$ changes differently with temperature, and $m$ is the effective (band) mass \cite{bardeen1940}.

Hence, the entropy production rate \eqref{eq:eprdelay} becomes
\begin{equation}
\dot{\Sigma} =\frac{\sigma_0}{\tau_{R}}\, E(t) \int_{0}^{t}dt'\,\e{-|t-t'|/\tau_{R}}\, E(t')\,,
\label{eq:eprDS}
\end{equation}
where $\sigma_0$ is the zero-frequency conductivity of the Drude-Sommerfeld model \cite{Ashcroft1976,Maslov2016}, and we assume $E(t)\neq 0$ for $t>0$ only.

Now further assuming the simplest parametrization for the electric field, $E(t)=E_0 \sin\left(\omega_0 t\right)$, and employing Eq.~\eqref{eq:eprDS} the entropy production rate \eqref{eq:eprdelay} becomes,
\begin{equation}
\begin{split}
&\dot{\Sigma} = \frac{\sigma_0\,E^{2}_{0}\,\sin{(\omega_{0} t)}}{1+(\omega_{0}\tau_{R})^{2}}\\
&\times \left[\omega_{0}\tau_{R}\,\e{-t/\tau_{R}}-\omega_{0}\tau_{R}\,\cos{(\omega_{0} t)}+\sin{(\omega_{0} t)}\right].
\end{split}
\label{eq:eprharm}
\end{equation}
In the Supplemental Material \cite{suppl} we provide a numerical illustration of this expression for $\dot{\Sigma}$. Obviously, Eq.~\eqref{eq:eprharm} can take marked negative values. Moreover, negative rates persist even when $\tau_{0}=2 \pi/\omega_0$ is orders of magnitude larger than $\tau_{R}$, which shows that this phenomenon is not restricted to microscopic time scales. For large periods of oscillation and $\omega_{0}\tau_{R}\to 0$, $\dot{\Sigma}$ \eqref{eq:eprharm} asymptotically becomes
\begin{equation}
\dot{\Sigma}\simeq \sigma_0 E_{0}^{2}\,\sin{(\omega_{0} t)}\left[-\omega_{0} \tau_{R}\, \cos{(\omega_{0} t)} + \sin{(\omega_{0} t)} \right]\,,
\end{equation}
which shows that there is always a small vicinity of $\omega_{0} t = n\pi$, $n=1, 2, 3,\ldots$, with $\dot{\Sigma} < 0$ as long as $\omega_{0} \tau_{R}\neq 0$.

The emergence of negative values of $\dot{\Sigma}$ can also be understood qualitatively. In Eq.~(\ref{eq:eprDS}) we observe that $\dot{\Sigma}$ is given by the product of a time-dependent electric field and a convolution between the same field (evaluated at a previous time) and the response function. This convolution describes a delay between the response of the system and the external driving. Thus, if the function $E(t)$ is non-monotonic and acquires negative values, this delay leads to negative values of $\dot{\Sigma}$. 

Now, the natural question arises whether the entropy production, $\Sigma=\int_{0}^{t}dt'\,\dot{\Sigma} $, itself remains positive at all times. To this end, we write
\begin{equation}
\label{eq:entprod}
\Sigma=\frac{1}{2}\int_{0}^{t}dt' \int_{0}^{t}dt''\,E(t') \phi(t'-t'') E(t'')
\end{equation}
which follows from simple manipulations of the integral and using $\phi(-t)=\phi(t)$. For such quadratic forms it was recently shown \cite{naze2019} that the entropy production is non-negative iff the Fourier transform of the response function, $\hat{\phi}(\omega) = \int dt\,\e{i\omega t} \phi(t)$, is a non-negative function.

\begin{figure}
	\includegraphics[width=0.48\textwidth]{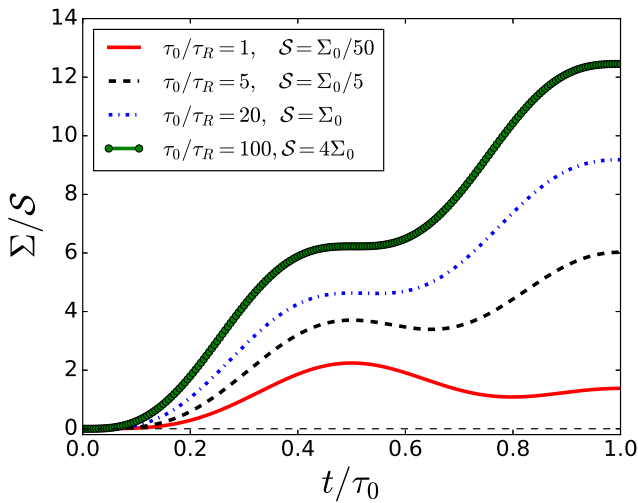}
	\caption{(Color online) Entropy production \eqref{eq:entprod} as a function of time for the Drude-Sommerfeld model under AC driving in units of $\Sigma_0 \equiv \sigma_0 E_0^2 \tau_R$. }
\label{fig:epharm}
\end{figure}

For the Drude-Sommerfeld model we obtain
\begin{equation}
\hat{\phi}(\omega) = \frac{2\, \sigma_0}{1+(\omega \tau_{R})^{2}}\,,
\label{eq:ft2respfunc}
\end{equation}
which is clearly positive. Thus, also the entropy production $\Sigma$ is a non-negative, albeit non-monotonic function of time. In Fig.~\ref{fig:epharm}  we depict $\Sigma$ for the entropy production rate \eqref{eq:eprharm} for a range of parameters. The plot clearly demonstrates the theoretical prediction. Curiously, comparing with Ref.~\cite{Williams2007} we also observe the stark thermodynamic similarity of the Drude-Sommerfeld model and viscoelastic fluids.

In the Supplemental Material \cite{suppl} we further show that the time-averaged entropy production rate is proportional to the Fourier transform $\hat{\phi}(\omega)$ of the response function. Thus, the non-negativity of $\hat{\phi}(\omega)$ might be understood simply as the standard statement of the fluctuation-dissipation theorem. However, we have also demonstrated that $\hat{\phi}(\omega)\geq 0$ is not equivalent to the positivity of $\dot{\Sigma}$. In other words, the second law of thermodynamics is \emph{not} synonymous to non-negative entropy production rates.

We also show in the Supplemental Material \cite{suppl} that negative rates become more salient as the frequency of the oscillatory field approaches $1/\tau_{R}$. However, it is known that the Drude-Sommerfeld model starts to deviate from experimental results for high frequencies \cite{Olmon2012,Yang2015}. Therefore, we continue the analysis with an extended model that accounts for more intricate quantum effects.

\paragraph{Extended Drude-Sommerfeld model}

The Extended Drude-Sommerfeld model is a phenomenological generalization accounting for the frequency dependence of the dielectric constant $\epsilon(\omega)$ \cite{Allen1977,Youn2007}. Specifically, the electrical conductivity is written as
\begin{equation}
\label{eq:EDScond}
\sigma(\omega)  = \frac{\epsilon_{0}\, \omega_{p}^{2}}{G(\omega) - i\omega}\,,
\end{equation}
where $\epsilon_{0}$ is the vacuum electrical permittivity, and $\omega_{p}$ is the plasma frequency with $\omega_{p}^{2} = N q^{2}/\epsilon_{0} m$. 

Comparing Eq.~\eqref{eq:classconduc} with Eq.~\eqref{eq:EDScond} we notice that the function $G(\omega)$ phenomenologically encodes the quantum many-body properties, and it is determined such that $\sigma(\omega)$ agrees with the Kramers-Kronig relations \cite{suppl}. Note that in the ``standard" Drude-Sommerfeld model we simply have $G(\omega)=1/\tau_{R}^0$, where $\tau_R^0$ now denotes the relaxation time at $\omega=0$. More generally, $G(\omega)$ is given by 
\begin{equation}
G(\omega) = \int_{0}^{\infty} dt\,\e{i\omega t}\,g(t)\,,
\end{equation}
with $g(t)$ real and $g(t) = 0$ for $t < 0$, which again is an expression of causality. It can be shown that $g(t)$ generalizes the collision term \eqref{eq:coll} of the Boltzmann equation to a convolution, which becomes local in time (i.e., memoryless) in the relaxation-time approximation \cite{Allen1977}.

Equation~\eqref{eq:EDScond} implies that the relaxation time becomes a function of frequency $\tau_{R}(\omega)$, and we have
\begin{equation}
\label{eq:tauomega}
\frac{1}{\tau_{R}(\omega)} = \frac{\omega \sigma_{1}(\omega)}{\sigma_{2}(\omega)} = \frac{G_{1}(\omega)}{1-G_{2}(\omega)/\omega}
\end{equation}
where the subscripts $1$ and $2$ denote real and imaginary parts, respectively. For noble metals, the electron-phonon interaction is fully covered by the relaxation time approximation, and the electron-electron interaction at the level of Fermi-liquid theory yields in leading order \cite{Beach1977,Maslov2016}
\begin{equation}
\frac{1}{\tau_{R}(\omega)} \approx \frac{1}{\tau_{R}^{0}} + b\,\omega^{2}\,,
\label{eq.tau1}
\end{equation}
where both $\tau_{R}^{0}$ and $b$ are functions of temperature.

To appropriately describe experimental findings, $G(\omega)$ must fulfill (i) that the order of magnitude of $\tau_{R}^{0}$ at room temperature is $\mc{O}(10^{-14})\,$s, and (ii) the relaxation time $\tau_{R}(\omega)$ has to match the experimentally measured behavior \eqref{eq.tau1}, with $b/\tau_{R}^{0}$ of the order of $\mc{O}(10^{-3})$ \cite{Yang2015}. Given these conditions, we make a phenomenological ansatz with three free parameters, $A$, $a$ and $\tau_{c}$,
\begin{equation}
\begin{split}
g(t) &= \Theta(t)\,A\left[\frac{1+(a\tau_{c})^{2}}{2\tau_{c}}\right]^{2} \e{-t/\tau_{c}}\\
&\quad\times\left[\cos{(a t)}+\sin{(a t)/a\tau_{c}}\right]\,.
\end{split}
\label{eq:g}
\end{equation}
where $\Theta(t)$ is again the Heaviside step function. %Simpler expressions could not account for all the experimental requirements described previously.
\begin{figure}
\centering
\includegraphics[width=0.48\textwidth]{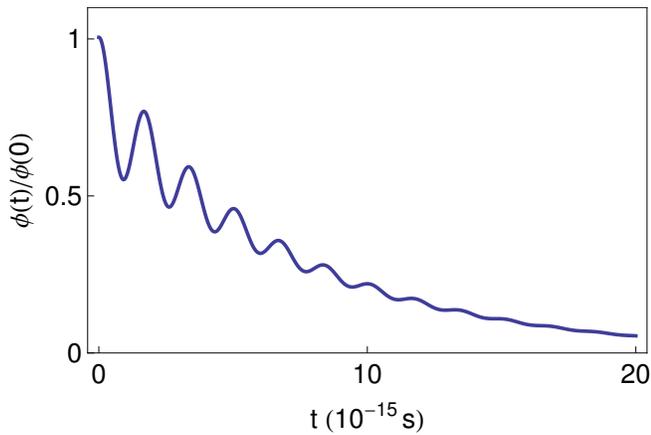}
\caption{\label{fig:response} (Color online) Response function for the extended Drude-Sommerfeld model resulting from the phenomenological ansatz \eqref{eq:g} and parameters: $A=10^{-2}$, $a\tau_{c}=10$ and $\tau_{c}=3\times10^{-15}\,$s.}
\end{figure}
In the Supplemental Material \cite{suppl} we compare our phenomenological description with the experimental findings in gold \cite{Olmon2012} and silver \cite{Yang2015}.  We find good agreement for conductivity in the ranges 0.1 to $\sim 4$ eV  of photon energies. 

Using the phenomenological ansatz \eqref{eq:g} we determine the conductivity $\sigma(\omega)$ \eqref{eq:EDScond}, and eventually the response function. The resulting expression is depicted in Fig.~\ref{fig:response} for experimentally relevant parameters. We observe that $\phi(t)$ is given by a sum of a simple exponential decay and another exponential decay (with a slightly faster rate) that multiplies an oscillatory function (see \cite{suppl} for details). 

\begin{figure}
\centering
\includegraphics[width=0.48\textwidth]{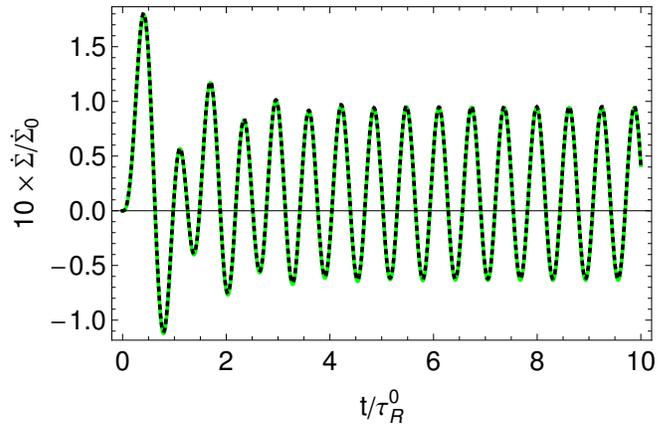}
\caption{\label{fig:EPR_EDS} (Color online) Entropy production rates, $\dot{\Sigma}$, in units of $\dot{\Sigma}_{0}=\sigma_{0}E_{0}^{2}$ for photon energy of 0.5 eV for the Drude-Sommerfeld model \eqref{eq:eprharm} (green solid line) and the Extended Drude-Sommerfeld model (black dotted line). Parameters are $A=10^{-2}$, $a\tau_{c}=10$ and $\tau_{c}=3\times10^{-15}\,$s.}
\end{figure}

In Fig.~\ref{fig:EPR_EDS} we plot the entropy production rate for the Drude-Sommerfeld model \eqref{eq:eprharm} together with the rate resulting for the Extended Drude-Sommerfeld model. Parameters are chosen to be experimentally relevant and for a photon energy of 0.5 eV. We observe (i) that the Drude-Sommerfeld model and the Extended Drude-Sommerfeld model result in barely distinguishable rates, (ii) the entropy production rate has marked negative periods, and (iii) after an initial transient the driven system relaxes into a periodic stationary state.

In conclusion, we have demonstrated that the entropy production rate in metals under AC driving can take negative values. This observation is not an artifact of the overly simplistic Drude and Drude-Sommerfeld model, but is also exhibited by the Extended Drude-Sommerfeld model that describes experimental data to high accuracy \cite{Olmon2012,Yang2015}.

\paragraph{Experimental verification and generalizations}

As alluded to above, the occurrence of negative values of the entropy production rate is governed by the competition of the rate of relaxation time and the rate of driving. Typically, $1/\tau^{0}_{R}$ varies from $10^{9}$ to $10^{14}$Hz depending on the temperature and the impurity concentration \cite{Olmon2012,Yang2015}. These rates are well within experimentally accessible driving. From a more practical point of view, negative entropy production rates are intimately related to negative heating rates \cite{Williams2007}. Thus, these negative rates might be directly measured in metals if it becomes possible to measure the temporally resolved, local temperature. However, nanothermometry still poses significant challenges with many open questions \cite{brites2012thermometry,Menges2016,Menges2016_2,Campbell2018}

In the present case, the entropy production rate has the form of current times electric field. We know from the theory of electrodynamics that this is exactly minus the power transferred from the electromagnetic radiation to the charges in the conductor. Thus, we have just described a situation in which the energy delivered by radiation is irreversibly absorbed by a thermally isolated system consisting of a collection of electrons plus its heat bath of phonons. However, we know from every-day life experience that metals reflect part of the visible light that we shine on them. This seems to be the experimental evidence that entropy production rates can become negative since they are a measure of absorbed power. 

Additionally, the physical mechanism leading to negative entropy production rates is not restricted to free electron gases. For instance, similar behavior of the response function can also be found in the Hubbard model \cite{karrasch2014}, which suggests that analogous studies and/or experiments could be performed in more complicated materials. In addition, conductivity measurements of ultracold atoms in optical lattices have already been performed and might constitute an alternative setup to bulk metals \cite{anderson2019}. 

Finally, experiments on levitated nanoparticles have been reported that also exhibit potentially negative entropy production rates \cite{Gieseler2018}. However, similarly to sheared viscoelastic fluids \cite{Williams2007}, levitated nanoparticles \cite{Gieseler2018} are somewhat farther from home than electrical conduction.

\paragraph{Concluding remarks}

In the present work, we have demonstrated that in the rather mundane situation of electric conduction under AC driving the entropy production rate assumes negative values. This is \emph{not} a peculiarity of a system driven far from equilibrium, but rather a consequence of a competition of time scales in the linear regime.  We have found the same behavior for three variants of the Drude model, ranging from the classical free electron gas to a phenomenological extension accounting for all underlying quantum many-body effects. Finally, we emphasize that the positivity of the entropy production rate is not synonymous to the second law of thermodynamics, and that negative values do not necessarily indicate the end of the universe. 

\acknowledgements{The authors would like to thank R. Pereira, E. Miranda, G. T. Landi, M. Hinczewski for insightful discussions and constructive suggestions. P. Naz\'e and M. V. S. Bonan\c{c}a acknowledge financial support from FAPESP (Funda\c{c}\~{a}o de Amparo \`a Pesquisa do Estado de S\~ao Paulo) (Brazil) (Grant No. 2018/06365-4 and No. 2018/21285-7). }

\newpage
\appendix

\section{Conductivity and Green-Kubo formula}

We briefly show here how to express the conductivity in terms of Green-Kubo formula. We start with the following Hamiltonian,
\begin{equation}
H(t) = H_{0} - q\sum_{k=1}^{N} \mathbf{r}_{k}\cdot \mathbf{E}(t)\,,
\label{eq:drudemodel}
\end{equation}
where $H_0$ is the bare Hamiltonian of the electron gas plus its heat bath, and $\mathbf{E}(t)$ is the time-dependent electric field. Moreover, $q$ is the charge and $\mathbf{r}_{k}$ denotes the position of the $k$th electron. We further assume that the material is isotropic and homogeneous, and only the magnitude of $\mathbf{E}(t)$ changes while its direction remains constant. In the simplest case, we have $\mathbf{E}(t) =\mathbf{E}_0 \sin\left(\omega_{0} t\right)$, but the following analysis remains true for more complicated driving.

The standard approach \cite{kubo2012,Ashcroft1976,giuliani2005} then allows to compute the electrical current, $\mathcal{J}(t)$, in terms of the response function. We have,
\begin{equation}
\mathcal{J}(t) = \int_{-\infty}^{t}dt'\,\Phi(t-t') E(t')\,,
\label{eq:noneqcurr}
\end{equation} 
where we dropped the vector notation, since the direction of the electric field is constant. The response function at inverse temperature $\beta$ is determined by \cite{kubo2012},
\begin{equation}
\Phi(t) = \Theta(t) \int_{0}^{\beta} d\lambda\,\langle J(-i\hbar\lambda) J(t)\rangle = \Theta(t)\, \phi(t)\,,
\label{eq:respfunc1}
\end{equation}
and where we included the Heaviside step function $\Theta(t)$ to preserve causality. Moreover, $J(t)$ is the total electric flux operator in the interaction picture,
\begin{equation}
J(t) = \e{i H_0  t/\hbar}\left(q \sum_k \dot{u}_{k}\right) \e{-i H_0 t/\hbar}\,,
\end{equation}
where  the $\dot{u}_{k}$ are the time derivatives of the projection of $\mathbf{r}_{k}$ in the direction of $\mathbf{E}(t)$. Finally, the flux correlation function in Eq.~\eqref{eq:respfunc1} reads
\begin{equation}
\begin{split}
\langle J(-i\hbar\lambda) J(t)\rangle =\frac{\tr{\e{-\beta H_{0}} J(-i\hbar\lambda) J(t)}}{\tr{\e{-\beta H_{0}}}}\,,
\end{split}
\end{equation}
which is a thermal average with respect to the bare Hamiltonian $H_0$. Hamiltonian dynamics demands that the response function in Eq.~\eqref{eq:respfunc1} fulfills $\phi(-t) = \phi(t)$ (see Ref.~\cite{kubo2012}), and is thus thermodynamically consistent \cite{naze2019}.

%%%%%%%%%%%%%%%%%%%%%%%%%%%%%%%%%%%%%%%%%%%%%%%%
\section{Non-negativity of entropy production}
\label{app:proof}

A question that naturally arises from our results is whether the entropy production $\Sigma$ obtained from the time integral of $\dot{\Sigma}$ is always non-negative. Mathematically, this issue can be addressed along the same lines of what was done in Ref.~\cite{naze2019}. First, we rewrite the time integral appropriately as,
\begin{equation}
\Sigma = \frac{1}{2}\int_{0}^{t}dt' \int_{0}^{t}dt''\,E(t')\, \phi(t'-t'')\, E(t'')\,,
\label{eq:epgeneral}
\end{equation} 
where the property $\phi(-t)=\phi(t)$ (see previous Section) was used. Afterwards, the following theorem applies.
\begin{theorem}
The entropy production $\Sigma$ given by Eq.~(\ref{eq:epgeneral}) is non-negative if, and only if, the Fourier transform $\hat{\phi}(\omega)$ of $\phi(t)$ is a non-negative function, i.e.
\begin{equation}
\hat{\phi}(\omega) = \int_{-\infty}^{\infty}dt\,\e{i\omega t}\, \phi(t) \geq 0\,.
\label{eq:ftrespfunc}
\end{equation}
\label{te:teo1}
\end{theorem}
We will restrict ourselves to prove the first implication of Theorem~\ref{te:teo1}. The second one can be seen in detail in Ref.~\cite{feller2008} under the name \emph{Bochner's theorem}. We start using the fact that $\hat{\phi}(\omega)$, as defined in Eq.~(\ref{eq:ftrespfunc}), is real since $\phi(-t) = \phi(t)$. Hence, if $\hat{\phi}(\omega)$ is non-negative, we can write Eq.~(\ref{eq:epgeneral}) as follows,
\begin{equation}
\begin{split}
\Sigma& = \frac{1}{4\pi}\int_{0}^{t}dt' \int_{0}^{t}dt''\,E(t')E(t'')\,\\
&\times\int_{-\infty}^{\infty}d\omega\,\e{-i\omega(t'-t'')}\,\hat{\phi}(\omega)\\
&= \frac{1}{4\pi}\int_{-\infty}^{\infty}d\omega\,\left|\int_{0}^{t}dt'\,\e{-i\omega t'}\, E(t') \right|^{2} \hat{\phi}(\omega)\geq 0\,,
\end{split}
\end{equation}
which proves the first implication. It is straightforward to further verify that the Fourier transform of $\phi(t) = (\sigma_{0}/\tau_{R})\e{-|t|/\tau_{R}}$,
\begin{equation}
\hat{\phi}(\omega) = \frac{2 \sigma_{0}}{\tau_{R}}\int_{0}^{\infty}dt\,\e{-t/\tau_{R}}\, \cos{(\omega t)} = \frac{2 \sigma_{0}}{1+(\omega \tau_{R})^{2}}\,,
\label{eq:ft2respfunc}
\end{equation}
is always non-negative. 

%%%%%%%%%%%%%%%%%%%%%%%%%%%%%%%%%%%%%%%%%%%%%%%%%%%%%
\section{Fluctuation-Dissipation Theorem}
\label{app:fdt}

We now that the time-averaged entropy production (TEP) rate is proportional to the Fourier transform of the response function $\phi(t)$. To this end, we define the TEP rate as follows,
\begin{equation}
\overline{\dot{\Sigma}} = \frac{1}{\tau}\int_{0}^{\tau}dt\,\dot{\Sigma} = \overline{E(t)\mathcal{J}(t)}\,,
\label{eq:tepr}
\end{equation}
where $\tau=2\pi/\omega$ is the period of oscillation of $E(t)$. Since the current $\mathcal{J}(t)$ is given by
\begin{equation}
\mathcal{J}(t) = \int_{-\infty}^{t}dt'\,\Phi(t-t') E(t') = \int_{0}^{\infty}dt'\,\Phi(t') E(t-t')\,,
\end{equation}
with $\Phi(t) = \Theta(t)\phi(t)$ and $E(t) = E_{0}\sin{(\omega t)}$, we obtain
\begin{equation}
\mathcal{J}(t) = \frac{E_{0}}{2 i}\left( e^{i\omega t}\chi^{*}(\omega) - e^{-i\omega t}\chi(\omega)\right)\,,
\end{equation}
where 
\begin{equation}
\chi(\omega) = \int_{0}^{\infty}dt\,e^{i\omega t}\phi(t) = \chi_{1}(\omega)+i \chi_{2}(\omega)\,.
\end{equation}

The current $\mathcal{J}(t)$ can be written then in terms of $\chi_{1}(\omega)$ and $\chi_{2}(\omega)$, the real and imaginary parts of $\chi(\omega)$, respectively,
\begin{equation}
\mathcal{J}(t) = E_{0}\left(\sin{(\omega t)} \chi_{1}(\omega) - \cos{(\omega t)} \chi_{2}(\omega)\right)\,,
\end{equation} 
that yields the following expression for TEP rate,
\begin{equation}
\overline{\dot{\Sigma}} = \frac{E_{0}^{2}}{\tau}\int_{0}^{\tau}dt\,\sin^{2}{(\omega t)}\chi_{1}(\omega) = \chi_{1}(\omega)\,\frac{E_{0}^{2}}{2}\,.
\label{eq:fdt}
\end{equation} 

Equation~(\ref{eq:ft2respfunc}) implies that $\chi_{1}(\omega) = \hat{\phi}(\omega)/2$, i.e. the TEP rate, which is a measure of the averaged dissipated power, is given in terms of the Fourier transform $\hat{\phi}(\omega)$ of the response function. Hence, $\hat{\phi}(\omega)$ must be non-negative if we demand that $\overline{\dot{\Sigma}}\geq 0$. Since Green-Kubo formula relates $\hat{\phi}(\omega)$ to the Fourier transform of certain equilibrium fluctuations, the non-negativity of Eq.~(\ref{eq:fdt}) becomes the standard statement of the fluctuation-dissipation theorem \cite{kubo2012}.

%%%%%%%%%%%%%%%%%%%%%%%%%%%%%%%%%%%%%%%%%%%%%%%%%%
\section{Drude-Sommerfeld model}

As discussed in the main text, the response function $\phi(t) = (\sigma_{0}/\tau_{R})\exp{(-|t|/\tau_{R})}$ of the Drude-Sommerfeld (DS) model leads to the following expression for entropy production rate (EPR) when $E(t) = E_{0}\sin{(\omega_{0}t)}$,
\begin{equation}
\begin{split}
&\dot{\Sigma} = \frac{\sigma_0\,E^{2}_{0}\,\sin{(\omega_{0} t)}}{1+(\omega_{0}\tau_{R})^{2}}\\
&\times \left[\omega_{0}\tau_{R}\,\e{-t/\tau_{R}}-\omega_{0}\tau_{R}\,\cos{(\omega_{0} t)}+\sin{(\omega_{0} t)}\right]\,.
\end{split}
\label{eq:eprharm}
\end{equation}

\begin{figure*}[t]
 \subfigure{\includegraphics[scale=0.42]{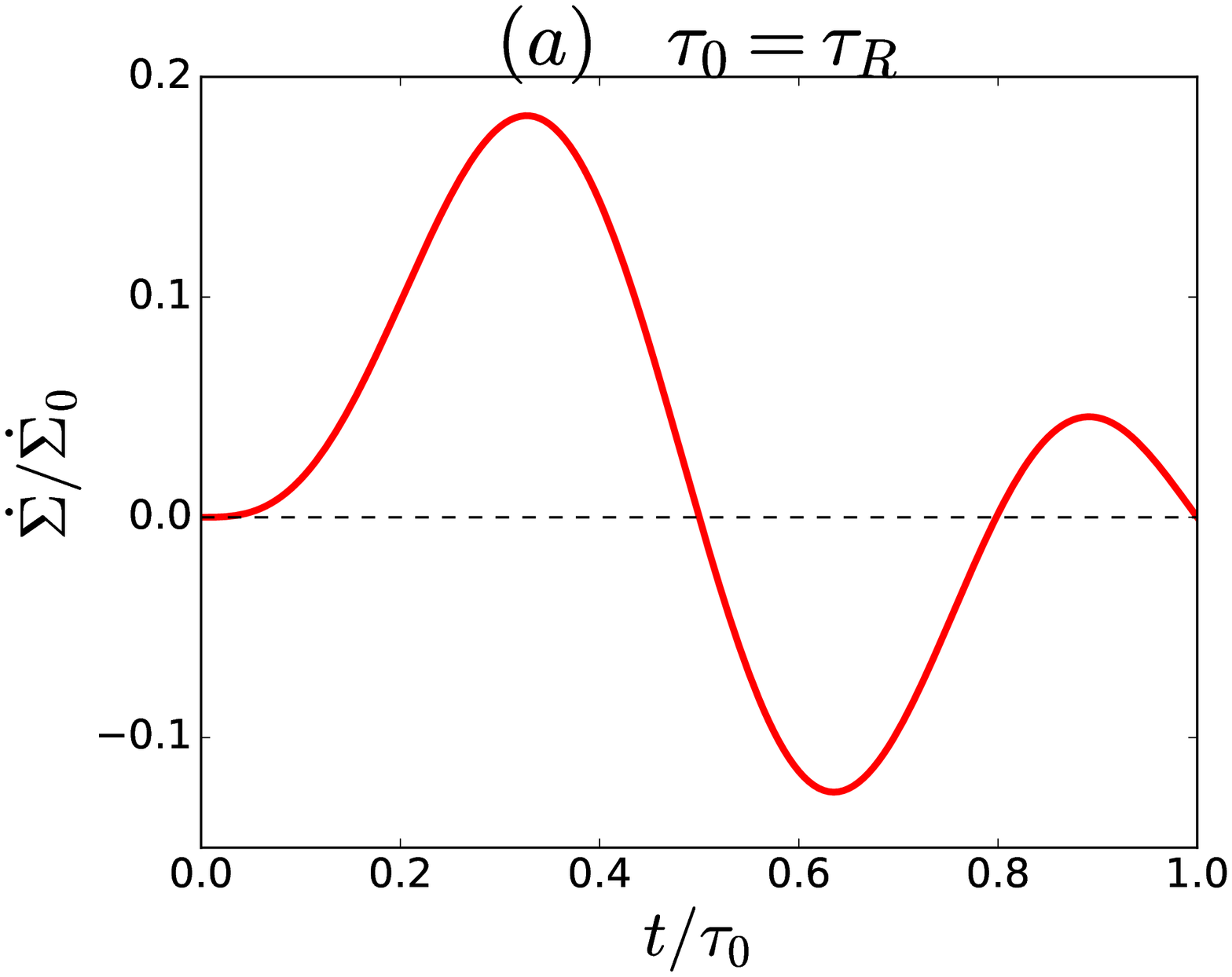}}
 \subfigure{\includegraphics[scale=0.42]{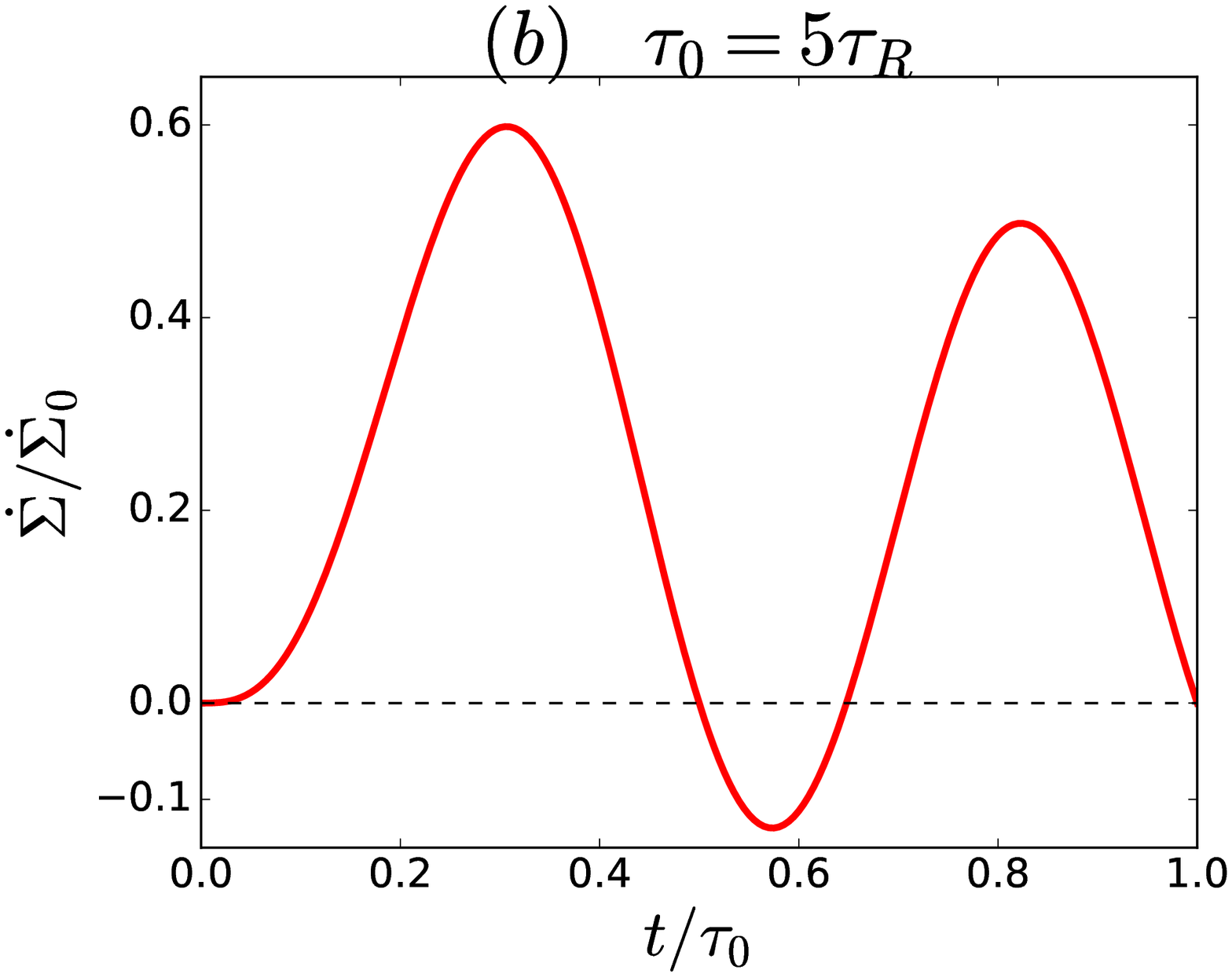}}\\
 \subfigure{\includegraphics[scale=0.42]{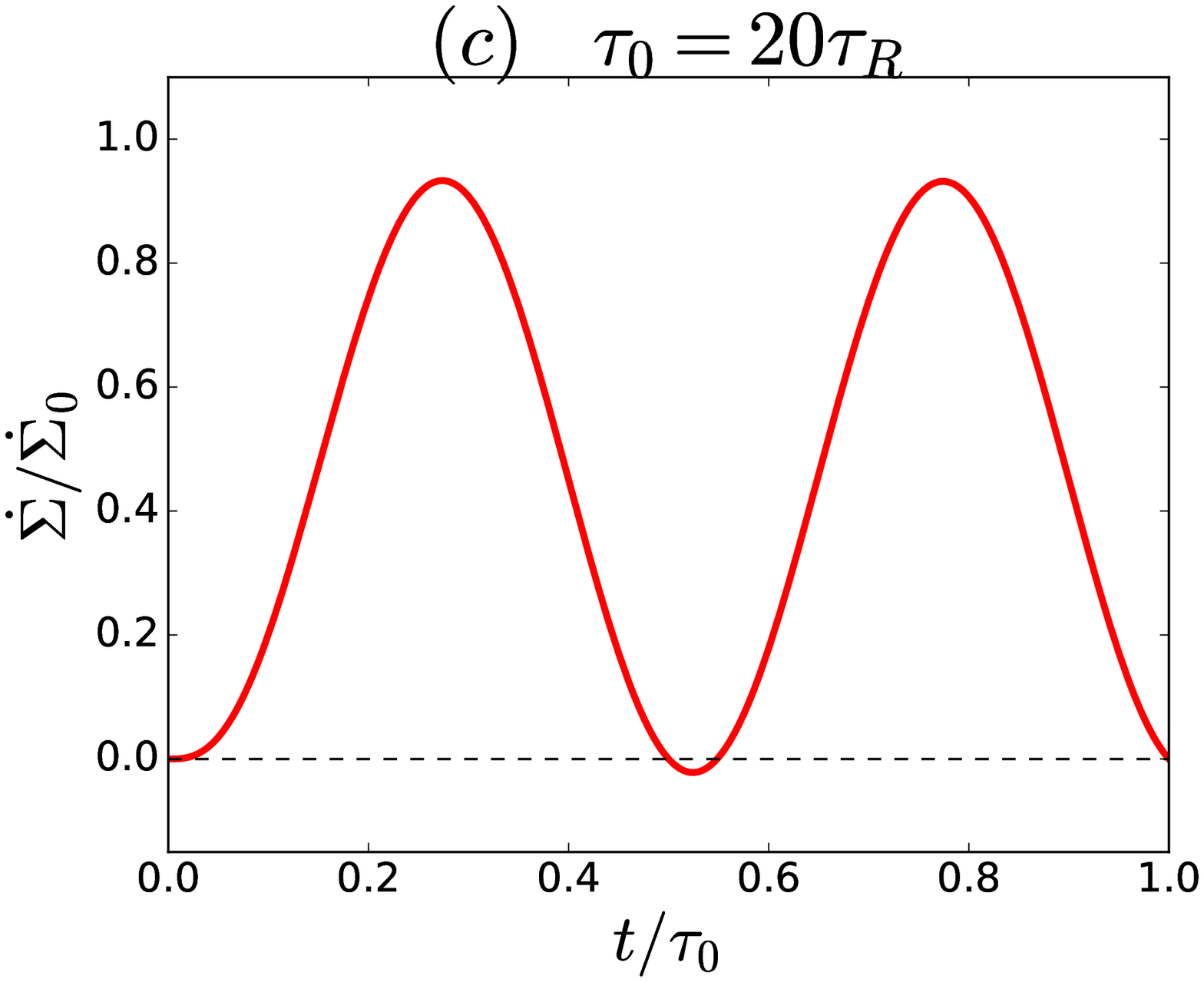}}
 \subfigure{\includegraphics[scale=0.42]{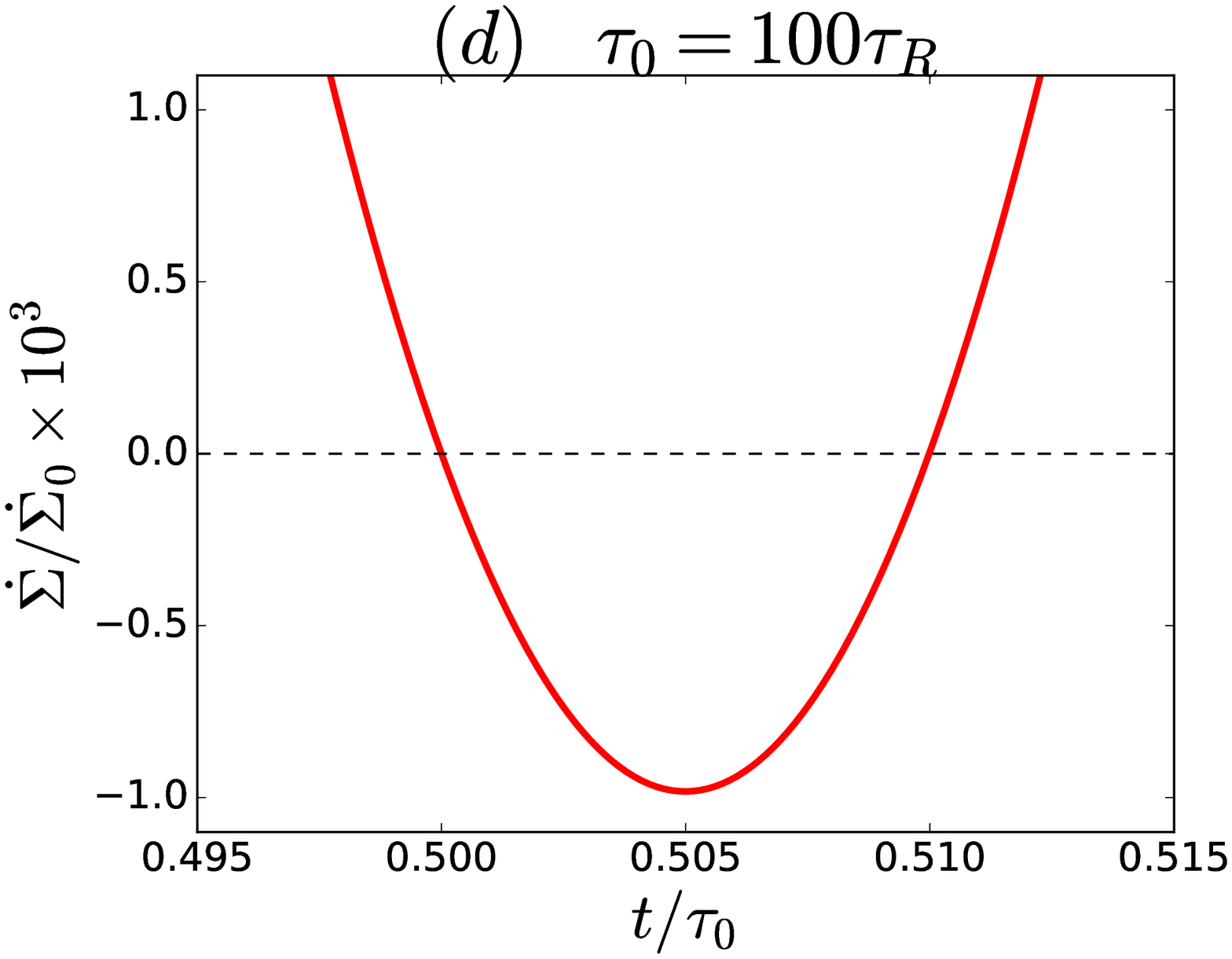}}
\caption{(Color online) Entropy production rates given by Eq.~(\ref{eq:eprharm}) for the following values of the ratio $\tau_{0}/\tau_{R}$: (a) 1, (b) 5, (c) 20 and (d) $10^{2}$. We defined $\dot{\Sigma}_0 = \sigma_{0} E_0^2$.}
\label{fig:eprharm}
\end{figure*}

Figure~\ref{fig:eprharm} shows the behavior of Eq.~(\ref{eq:eprharm}) for different values of $\tau_{0}/\tau_{R}$ ($\tau_{0}=2\pi/\omega_{0}$). Notice that the magnitude of the negative values of $\dot{\Sigma}$ decrease as $\tau_{0}/\tau_{R}$ increases.

%%%%%%%%%%%%%%%%%%%%%%%%%%%%%%%%%%%%%%%%%%%%%%%%%%%%%%%%%%%%%%%%%%%%
\section{Extended Drude-Sommerfeld model}

Our phenomenological proposal of an extended Drude-Sommerfeld model (EDS) relies on the following expression for the electrical conductivity \cite{Allen1977,Youn2007},
\begin{equation}
\label{eq:EDScond}
\sigma(\omega)  = \frac{\epsilon_{0}\, \omega_{p}^{2}}{G(\omega) - i\omega}\,,
\end{equation}
where $\epsilon_{0}$ is the vacuum electrical permittivity, and $\omega_{p}$ is the plasma frequency. For $G(\omega) = 1/\tau_{R}$, we recover the DS model. Thus, the function $G(\omega)$ introduces a frequency-dependent relaxation time whose underlying physical mechanism is the electron-electron interaction \cite{Olmon2012,Yang2015,Maslov2016}.

We demand that Eq.~(\ref{eq:EDScond}) fulfills the Kramers-Kronig or dispersion relations (which are an expression of causality \cite{nussenzveig}). For that, $\sigma(\omega)$ must be analytic in the upper-half complex plane, i.e., poles of $\sigma(\omega)$ must have negative imaginary parts (for details, see Titchmarsh theorem \cite{nussenzveig}). Additionally, the response function obtained from Eq.~(\ref{eq:EDScond}) through
\begin{equation}
\Phi(t) = \frac{1}{2\pi} \int_{-\infty}^{\infty} d\omega\,e^{-i\omega t}\sigma(\omega)\,,
\label{eq:respfunc}
\end{equation}
must be real. This requires that $\sigma_{1}(-\omega) = \sigma_{1}(\omega)$ and $\sigma_{2}(-\omega) = -\sigma_{2}(\omega)$, where $\sigma_{1}(\omega)$ e $\sigma_{2}(\omega)$ denote the real and imaginary part of $\sigma(\omega)$, respectively.

All the requirements mentioned above can be achieved by a suitable choice of $G(\omega)$. First, simple algebra shows that $G_{1}(-\omega) = G_{1}(\omega)$ and $G_{2}(-\omega) = -G_{2}(\omega)$ (where subscripts 1 and 2 denote real and imaginary parts of $G(\omega)$, respectively) imply the correct parity of $\sigma(\omega)$. Second, by choosing $G(\omega)$ analytic in the upper-half complex plane, we will verify in the next section that $\sigma(\omega)$ is also analytic in the upper-half complex plane and hence satisfies Kramers-Kronig relations. In summary, $G(\omega)$ can be expressed as
\begin{equation}
G(\omega) = \int_{0}^{\infty}dt\,\e{i\omega t}\, g(t)\,,
\label{eq.gfunc}
\end{equation}
where the function $g(t)$ is real and $g(t) = 0$ for $t < 0$.

To give a hint about the physical meaning of $g(t)$, we recall that transport coefficients are often calculated using the Boltzmann equation \cite{Ashcroft1976}. In this approach, DS conductivity is obtained in the so-called relaxation-time approximation, which assumes that the collision term of Boltzmann equation can be written as $(\partial f/\partial t)_{\mathrm{coll}} = -\delta f/\tau_{R}^{0}$, where $f$, $\delta f = f - f_{0}$ and $\tau_{R}^{0}$ denote, respectively, the non-equilibrium distribution, the deviation from the initial equilibrium distribution $f_{0}$ (which in the present case is the Fermi-Dirac one) and the relaxation time. For the EDS, the collision term is written as \cite{Allen1977}  
\begin{equation}
\left( \frac{\partial f}{\partial t} \right)_{\mathrm{coll}} = -\int_{-\infty}^{t}dt'\,g(t-t')\,\delta f(t')\,,
\label{eq.collision}
\end{equation}
where $g(t)$ is the function introduced in Eq.~(\ref{eq.gfunc}). If $g(t)$ decays sufficiently fast, $\delta f$ can be taken outside the integral in Eq.~(\ref{eq.collision}) and the relaxation-time approximation is recovered.

As stated above, our approach does not provide a microscopic derivation of $g(t)$. In what follows, we show that an expression of $g(t)$ in terms of elementary functions accounts for a minimum list of experimental facts. Our results are based on experiments compiled in Refs.~\cite{Olmon2012,Yang2015} for the optical dielectric function of Au and Ag. These experiments show that DS model describes very well the behavior of these two noble metals in the range of  0.1 to 1.5 eV. EDS is understood then as a correction of DS model for photon energies larger than 1.5 eV. Additionally, Refs.~\cite{Olmon2012,Yang2015} show that
\begin{itemize}
\item The zero-frequency relaxation time $\tau_{R}^{0}$ must be of the order of $10^{-14}\,$s at room temperature;
\item The frequency-dependent relaxation time $\tau_{R}(\omega)$ must have the following low-frequency behavior (which is also predicted by Fermi-liquid theory \cite{Maslov2016}),
\begin{equation}
1/\tau_{R}(\omega) = \omega\sigma_{1}(\omega)/\sigma_{2}(\omega) \approx 1/\tau_{R}^{0} + b \omega^{2}\,,
\end{equation}
with the ratio $b/\tau_{R}^{0} \sim 10^{-3}$ at room temperature;
\end{itemize}
\begin{figure}
\centering
\includegraphics[scale=0.63]{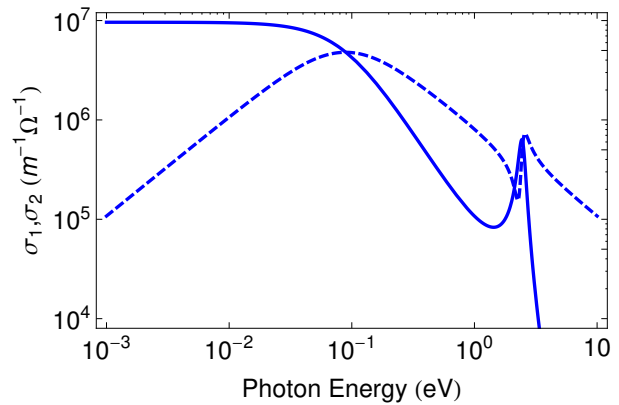}
\caption{(Color online) Real (solid line) and imaginary (dashed line) parts of the conductivity $\sigma(\omega)$ obtained from Eqs.~(\ref{eq:EDScond}), (\ref{eq.gfunc}) and (\ref{eq.gfinal}) with the following choice of parameters: $A=10^{-2}$, $a\tau_{c}=10$ and $\tau_{c}=3\times10^{-15}\,$s. The value of $\hbar\omega_{p} = 8.9$ eV for the plasma frequency was taken from the experimental results reported in Refs.~\cite{Olmon2012,Yang2015}.} 
\label{fig:conduc}
\end{figure}

Figure~\ref{fig:conduc} shows the result obtained for the conductivity given by Eq.~(\ref{eq:EDScond}) with the following choice of $g(t)$,
\begin{equation}
\begin{split}
&g(t) =\Theta(t)\,\left( \frac{2\tau_{c}}{1+(a\tau_{c})^{2}}\right)^{-2}\\
 &\quad\times A \e{-t/\tau_{c}}\left[ \cos{(a t)} + (1/(a\tau_{c}))\sin{(a t)}\right]\,,
\label{eq.gfinal}
\end{split}
\end{equation}
where $\Theta(t)$ is the Heaviside step function and $A$, $a$ and $\tau_{c}$ are free parameters whose values are determined by the above mentioned experimental facts. Our results can be directly compared to the conductivity of Au presented in Fig.~8 of Ref.~\cite{Olmon2012} and of Ag presented in Fig.~7 of Ref.~\cite{Yang2015}. They show a reasonable agreement with experiments and provide $\tau_{R}^{0} \approx 6\times 10^{-15}$s (roughly 3 times smaller than the relaxation time of Ag and Au \cite{Olmon2012,Yang2015}) and $b/\tau_{R}^{0} \approx 3\times 10^{-3}$ for the following choice of parameters: $A=10^{-2}$, $a\tau_{c}=10$ and $\tau_{c}=3\times10^{-15}\,$s, and the experimental value of $\hbar\omega_{p} = 8.9$ eV. 

\begin{figure}
\centering
\includegraphics[scale=0.63]{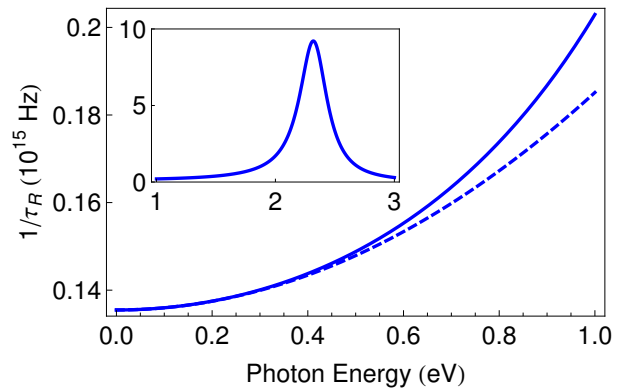}
\caption{(Color online) Low-frequency behavior of $1/\tau_{R}(\omega) = \omega\sigma_{1}(\omega)/\sigma_{2}(\omega)$ (solid lines) as predicted by our phenomenological EDS model (see Eqs.~(\ref{eq:EDScond}), (\ref{eq.gfunc}) and (\ref{eq.gfinal})) for the following choice of parameters: $A=10^{-2}$, $a\tau_{c}=10$ and $\tau_{c}=3\times10^{-15}\,$s, and the experimental value of $\hbar\omega_{p} = 8.9$ eV. The dashed line shows $1/\tau_{R}^{0} + b\omega^{2}$, with $\tau_{R}^{0}\approx 6\times 10^{-15}\,$s and $b/\tau_{R}^{0}\approx 3\times 10^{-3}$ obtained from our model. The inset shows that our model breaks down for energies between 2 and 3 eV.}
\label{fig:relax}
\end{figure}

Figure~\ref{fig:relax} shows that our EDS model provides a quadratic behavior for $1/\tau_{R}(\omega)$ at low frequencies. However, experiments show that $1/\tau_{R}(\omega)$ grows monotonically (and faster than $\omega^{2}$) up to photon energies of 5 to 10 eV \cite{Yang2015}. Our model fails to predict that already for photon energies between 2 and 3 eV.   

A few words about our choice of Eq.~(\ref{eq.gfinal}). The function $g(t)$ must decay sufficiently fast since the relaxation-time approximation must be recovered in some limit. We have verified that a simple exponential decay such as $A\exp{(-t/\tau_{c})}$ does not allow us to describe the minimum amount of experimental results described previously. Thus, Eq.~(\ref{eq.gfinal}) is the simplest function we have been able to find that furnishes a reasonably good agreement with experiments.

%%%%%%%%%%%%%%%%%%%%%%%%%%%%%%%%%%%%%%%%%%%%%%%%%%%%%%%%%%%%%%%%%%%
\paragraph*{Response function of the EDS model}

\begin{figure}
\centering
\includegraphics[scale=0.63]{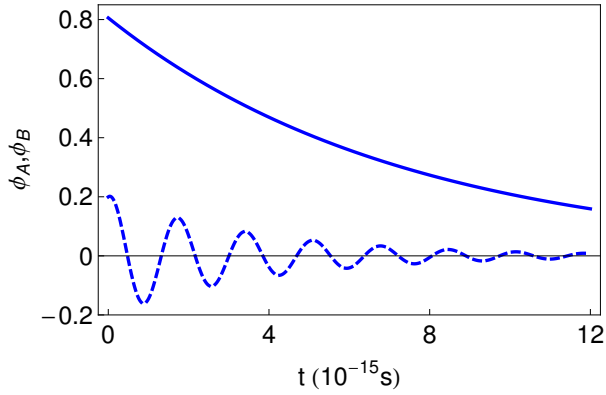}
\caption{(Color online) Time evolution of the contributions $\phi_{A}(t)$ (solid line) and $\phi_{B}(t)$ (dashed line) of the response function (\ref{eq:respfinal}) for the parameters $A=10^{-2}$, $a\tau_{c}=10$ and $\tau_{c}=3\times10^{-15}\,$s.}
\label{fig:respfunc} 
\end{figure}

Equations~(\ref{eq:EDScond}), (\ref{eq.gfunc}) and (\ref{eq.gfinal}) imply that
\begin{equation}
\sigma(\omega) = -i  \frac{\sigma_{0}}{f(\omega)}\frac{\tau_{c}}{\tau_{R}^{0}}\,[(a\tau_{c})^{2}-(i+\omega\tau_{c})]\,
\label{eq:EDSconduc2}
\end{equation}
where $\sigma_{0} = \epsilon_{0}\omega^{2}_{p}\tau_{R}^{0}$ and
\begin{equation}
\begin{split}
&f(\omega) = (\omega\tau_{c})^{3}+2i(\omega\tau_{c})^{2}-\frac{iA}{2}\,[1+(a\tau_{c})^{2}]^{2}\\
& \quad-\frac{\omega\tau_{c}}{4}\,[1+(a\tau_{c})^{2}][4+A(1+(a\tau_{c})^{2})]\,.
\end{split}
\end{equation}
The response function (\ref{eq:respfunc}) can be calculated then by contour integration in the complex plane once the poles of Eq.~(\ref{eq:EDSconduc2}) are obtained for the choice $A=10^{-2}$, $a\tau_{c}=10$ and $\tau_{c}=3\times10^{-15}\,$s. It can be shown that the poles are of the following form: $\omega_{1} = -i\alpha$, $\omega_{2} = \gamma - i\eta$ and $\omega_{3} = -\gamma - i\eta$, with $\alpha$, $\gamma$ and $\eta$ all positive. Thus, Eq.~\eqref{eq:EDSconduc2} fulfills Kramers-Kronig relations and the response function is expressed as the sum of two contributions,
\begin{equation}
\Phi(t) = \Theta(t) \frac{\sigma_{0}}{\tau_{R}^{0}}\left[\phi_{A}(t) + \phi_{B}(t) \right]\,,
\label{eq:respfinal}
\end{equation}
such that, for $t > 0$,
\begin{eqnarray}
\phi_{A}(t) &=& \phi_{A}(0)\, \e{-\alpha t}\,,\\
\phi_{B}(t) &=& \phi_{B}(0)\,\e{-\eta t}\,\left[\cos{(\gamma t)} + \kappa\sin{(\gamma t)}\right]\,,
\end{eqnarray}
where $\phi_{A}(0)$, $\phi_{B}(0)$, $\kappa$, $\alpha$, $\gamma$ and $\eta$ are constants determined by the values of $A$, $a\tau_{c}$ and $\tau_{c}$. Figure~\ref{fig:respfunc} shows the behavior of $\phi_{A}(t)$ and $\phi_{B}(t)$. Expression~\eqref{eq:respfinal} was used to obtain the entropy production rate shown in Fig.~3 of the main text.

%%%%%%%%%%%%%%%%%%%%%%%%%%%%%%%%%%%

\bibliography{ref1.bib}

\end{document}